# Thermally assisted domain wall nucleation in perpendicular anisotropy trilayer nanowires

Kulothungasagaran Narayanapillai, Xuepeng Qiu, Jan Rhensius, and Hyunsoo Yang[a]

*Department of Electrical and Computer Engineering, National University of Singapore, 117576, Singapore*

[a] e-mail address: eleyang@nus.edu.sg

We study thermally assisted domain wall generation in perpendicular magnetic anisotropy CoFeB trilayer nanowires by the effect of Joule heating. The anomalous Hall effect is utilized to detect magnetization reversal in order to study the domain wall generation. We observe a statistical distribution in the switching process which is consistent with the thermal activation process. Our results show that the proposed method provides an efficient way for generating domain walls in perpendicular magnetic nanowires at predefined locations.

# 1. Introduction

Magnetic domain walls (DWs) have attracted considerable attention of researchers due to the potential applications in storage[1] and logic devices[2]. DWs in perpendicular magnetic anisotropy (PMA) systems have advantages compared to the in-plane anisotropy systems. In particular, the DWs in PMA systems are narrow, leading to higher storage density, for example, 4.2 nm in Pt/Co/AlOx system[3], compared to up to a few hundred nanometers in the in-plane (permalloy) system[4]. Recent developments in PMA trilayer systems[5, 6] show their thin layer thicknesses can reduce the operation current for DW based devices. In addition, critical current density for current induced DW motion can be reduced by tuning the anisotropy through layer stacking[7]. Moreover, the spin Hall effect and the Rashba effect in these systems offer additional ability to manipulate the spin orbit induced torques[3, 8].

Effective DW generation and DW control plays an important role in realizing modern DW based devices. A few methodologies are used for DW generation in perpendicular systems. Utilizing the Oersted field is a common approach where a conductor is placed across a magnetic layer, which can be reversed by the Oersted field generated by a current pulse through the conductor[9]. Random DW generation by using a large magnetic reservoir is another popular approach where the reversal process depends on randomly distributed pinning sites on a large magnetic pad to nucleate DWs in an attached wire[10]. Other proposed methods include controlling the anisotropy by ion-irradiation[11] or ion milling[12] where the change/gradient in magnetic anisotropy is used to generate a DW. These methods introduce physical damages or locally vary the strength of the pinning sites which are not preferred in some device applications.

Energy assisted magnetization reversal has been studied extensively for applications in magnetoresistive random access memory (MRAM), spin valves, and bit pattern media where external energy is supplied to the system in the form of heat[13-15] or microwaves[16]. Thermal effects could be also utilized to reduce the pinning field[17] and the critical current density in current induced DW motion[18]. A recent study of energy assisted DW generation demonstrates the energy injection in the form of microwaves[19] into the nanowire, hence reducing the nucleation field.

In this letter, we study thermally assisted DW generation as an alternative method by introducing Joule heating in a well-defined portion of the nanowire. The applied current pulse amplitude to generate a DW shows a distribution. The required current amplitude increases, when the assist field to generate a DW decreases. Our method reduces one lithography step compared to the Oersted field generation process, while offering a reliable process to selectively generate DWs at predetermined locations with greater control.

## 2. Experimental results and discussion

Trilayer films exhibiting strong PMA with stacks of Pt (2 nm)/$Co_{60}Fe_{20}B_{20}$ (0.8 nm)/MgO (2 nm) (sample A), MgO (2 nm)/$Co_{60}Fe_{20}B_{20}$ (1.4 nm)/Ta (4 nm)/MgO (4 nm) (sample B), and Pt (2 nm)/$Co_{60}Fe_{20}B_{20}$ (0.8 nm)/MgO (2 nm)/$SiO_2$ (4 nm) (sample C) are deposited by magnetron sputtering with a base pressure of $2\times10^{-9}$ Torr on a Si substrate with a 100 nm thick thermally grown $SiO_2$ layer. Sample A and B have effective perpendicular anisotropy fields ($H_\perp^{eff}$) of 5.2 kOe and 5.6 kOe, respectively, at 200 K, while sample C shows 9.6 kOe at 300 K. A schematic of the layer stack structure for sample A is shown in Fig. 1(a). The films are post annealed at 250 °C for 1 hour. Figure 1(b) shows a scanning electron micrograph (SEM) of a device structure with a 600 nm wide nanowire with the same width as the Hall bars. The nanowires are patterned by electron beam lithography followed by Ar ion milling. The Hall bars are placed 8 μm apart. Contact electrodes are patterned by photolithography followed by Ta (5 nm)/Cu (105 nm) deposition by sputtering. The oxide layers are removed by Ar ion milling before depositing the contact pads to improve the electrical conductivity.

A DC current source is connected to $X_1X_2$ and voltmeters are connected across three Hall bars ($A_1A_2$, $B_1B_2$, and $C_1C_2$) as shown in Fig. 1(b) to locally probe the state of magnetization utilizing the anomalous Hall effect (AHE)[10]. The pulse generator is connected across $X_1B_1$, while the nanovoltmeter is connected to $A_1A_2$ for current pulsing experiments. The magnetic field is applied in the out-of-plane of the film direction (z-direction). The measured Hall resistances of $VH_1$, $VH_2$, and $VH_3$ are shown in Fig. 1(c) with an input DC current of 10 μA at 6 K. The nanowire shows coercivity ($H_C$) of 2445 Oe. The sharp switching of the Hall signals with the same switching field shows that there is no DW formation or strong pinning during the switching process. In all three Hall bars, the two magnetization states show a change of ~8 Ω. The small offset in the Hall resistance arises from the asymmetry of the patterned Hall bars. The temperature dependency of $H_C$ of the device is shown in Fig. 1(d). The coercivity decreases with increasing temperature. This strong temperature dependence of $H_C$ can be exploited for thermally assisted DW generation.

A schematic diagram to generate thermally assisted DWs is shown in Fig. 2(a) and the expected signals are sketched in Fig. 2(b). A Keithley 6221/2182A combination is used to apply current pulses and to measure the Hall voltage simultaneously. First, the device is saturated by applying a strong magnetic field $H_{SAT}$ = 4 kOe along the +z direction and back to zero field. The state of magnetization is probed by measuring the Hall signal across 3 Hall bars ($VH_1$, $VH_2$, and $VH_3$). All three readings initially show a high state (magnetization in the +z direction) as shown in Fig. 2(b). Then, a constant assist field $H_{ASSIST}$, which is smaller than the $H_C$ in order not to switch the magnetization, is applied along the –z direction. While keeping the field constant at $H_{ASSIST}$, the current is swept with increasing pulse amplitudes across $X_1B_1$ to induce local Joule heating in a part of the nanowire highlighted in red in Fig. 2(a). Due to an increase of the temperature in this nanowire portion, the local coercivity reduces. When the reduced coercivity equals to the external field, $H_{ASSIST}$, the magnetization in the heated segment reverses. $VH_1$ and $VH_2$ switches to a low state (–z direction), while $VH_3$ remains in the high state (+z direction), thereby generating a DW at the junction of the second Hall bar. Finally, the device is saturated in the opposite direction by applying $-H_{SAT}$ (–4 kOe) in order to remove any remanence and reset to the opposite state.

The distribution of the switching process is studied by repeatedly observing the switching behavior, employing the above described procedure. Current pulses with a pulse width of 1 ms and a pulse repetition interval of 5 s are applied across $X_1B_1$, while monitoring the Hall resistance $VH_1$. The pulse interval of 5 s is sufficient to bring the wire temperature back to pre-pulse state. Figure 3(a) shows the magnetization switching at 6 K from a low- to high-state with $H_{ASSIST}$ = 2037.5 Oe along the z-direction. Figure 3(b) shows the opposite sequence of a high- to low-state with $H_{ASSIST}$ = -2037.5 Oe at the same temperature. The current amplitude is swept from 200 µA to 350 µA and the switching is clearly represented by the abrupt change of the Hall resistance. In both sequences, we can see a switching distribution, since it involves a thermal activation process. The histograms of the switching events are presented in Fig. 3(c) and 3(d), respectively for each sequence. We further analyze the statistical distribution of the switching process. The switching distribution has a mean value (standard deviation) of 296.6 (27.7) µA and 294 (20.8) µA for each sequence, respectively. In our experimental geometry, 296.6 µA corresponds to a current density of $1.765 \times 10^{11}$ A/m$^2$. From the switching experiments, we can assume that the coercivity of the device should be reduced from 2445 ($H_C$) to 2037.5 Oe ($H_{ASSIST}$) by Joule heating. The corresponding effective temperature increase required to reduce the coercivity is found to be ~11 K from Fig. 1(d). The cumulative probability of switching is also shown in Fig. 3(c) and 3(d), respectively for each case. Above a certain amplitude of current pulse, the probability reaches to 100%.

We have further studied the DW generation process with varying $H_{ASSIST}$. While decreasing $H_{ASSIST}$, the required mean switching amplitude increases as shown in Fig. 3(e). This shows that we need to apply more energy to the nanowire in order to compensate the low value of $H_{ASSIST}$. In order to show that the heat assisted reversal process is dominated by Joule heating, we examine the effect of positive and negative currents on the switching process. The switching distribution is almost the same for both cases as seen in Fig. 3(e), since Joule heating does not depend on the current polarity. These results clearly show that a DW can be generated with the assistance of Joule heating. The switching current amplitude for the case of 50 μs pulse width is shown in Fig. 3(f) and the respective current density is plotted on the right y-axis. The required current density increases, when the current pulse width is changed from 1 ms to 50 μs. With a proper selection of the assist field and the pulse width, we can effectively control the DW generation process.

We have extended our studies to another trilayer system where a CoFeB layer is sandwiched between MgO and Ta (sample B) to verify the applicability of the proposed method. Similar device geometry and measurement methodology are utilized as discussed for sample A. The Hall resistance of the nanowire is shown in the inset of Fig. 4(a) at 6 K. The two magnetization states show a change of ~29 Ω across the Hall bar. The temperature dependent coercivity of the nanowire is plotted in Fig. 4(a). The required amplitude of switching current for DW generation with varying $H_{ASSIST}$ is shown in Fig. 4(b) at 6 K, which is similar to the case of sample A. In this experiment, a pulse width of 1 ms with a pulse repetition interval of 5 s is used for the experiments as well.

In order to further understand the thermal switching process, we have studied the DW generation process with respect to the pulse width in sample C. The stack structure is engineered to have a coercivity at room temperature with the addition of a capping material (4 nm $SiO_2$). The temperature versus coercivity field plot in Fig. 5 (a) shows that the coercivity of the sample C increases from 40 Oe to 205 Oe as the device is cooled from 300 K to 200 K. The switching amplitude is determined for every pulse width, ranging from 100 μs to 12 ms for a constant assist field of 25, 50, and 75 Oe at 200 K, respectively. The experimental setup limits the range of the pulse width applied in these measurements. A shorter pulse width requires a higher switching current density, while a longer pulse requires a lower switching current density. A higher $H_{ASSIST}$ on the other hand reduces the required switching current density and thus the required temperature change.

In the thermally activated regime, $\tau$ follows the Arrhenius-law based exponential equation, $\tau = \tau_0 \exp(E/K_B T)$ where $1/\tau_0$ is the attempt frequency ($\tau_0 = 10^{-11}$ s which are common for the perpendicular magnetic anisotropy) and $E$ is the activation energy. The activation energy can be described by $E = E_0 (1 - H_C / H_{th})^n$, in which the assist field is aligned along the easy axis of magnetization and opposing the initial magnetization. Here, $H_C$ is the coercive field which equals the assist field in our experiments and $H_{th}$ is the threshold field. The value of $n = 2$ is chosen to describe the thermally activated process of nucleation dominant magnetization reversal[20]. The rise time to reach the thermal equilibrium state is small (~2 μs) [21] compared to the applied pulse width (50 μs – 12 ms) in our experiments. Therefore, the temperature can be assumed to be constant during the pulse time. The reversal time is given by

$$\tau = \tau_0 \exp\left\{(1-H_C/H_{th})^n \frac{K_U V}{K_B T}\right\}$$

with the activation energy $E_0 = K_U V$ [22, 23]. In our experiments, the reversal time is equal to the pulse duration. An increase of temperature due to the effect of Joule heating in a nanowire is given by $\Delta T \propto J^2$, where $J$ is the current density [24, 25]. By rearranging the equation and incorporating the Joule heating effect, we obtain

$$\tau = \tau_0 \exp\left\{(1-H_C/H_{th})^n \frac{\beta}{\left(1+\frac{\alpha J^2}{T}\right)}\right\}, \text{ where } \beta\left(=\frac{K_U V}{K_B T}\right)$$

is the thermal stability factor. Fits of this equation with the experimental results are shown in Fig. 5(b). By rearranging the equation, we can describe the temperature dependence of coercivity as $H_C = H_{th}(1-aT^{1/n})$. By fitting with the experimental data in Fig. 5(a), we obtain the value of 1030 Oe for $H_{th}$. The extracted values of $\beta$ from the fittings in Fig. 5(b) are 51.4, 53, and 57.6 for 75, 50, and 25 Oe, respectively. $\beta$ has been reported to be 30 – 50 in a similar system[26]. Since $K_U$ increases with decreasing temperature in CoFeB/MgO systems[27], $\beta$ increases in our case measured at 200 K. Assuming a $K_U$ of $10^6$ erg/cm$^3$ with an extracted value of $\beta$ = 53, an activation volume (V) of $1.46 \times 10^{-24}$ m$^3$ is obtained. For our nanowire width of 600 nm and CoFeB thickness of 0.8 nm, the domain wall width is estimated to be 3.04 nm, which is similar to the previous report (4.2 nm) [3].

To examine the applicability of this method over a temperature range including room temperature, we have studied the DW depinning strength from the DW generated at different temperatures by the proposed heat assisted method with an assist field of 12 Oe. Figure 5(c) shows a DW depinning process at 260 K. When the DW leaves the Hall bar, a clear switching is observed in the Hall signals. The depinning field depends on the strength of the local pinning site and the coercivity of the material. The depinning field versus temperature for the range of 260 to 300 K is shown in Fig. 5(d) and the coercive field of the nanowire at respective temperatures is plotted as well. We can observe that the depinning field is always lower than the coercivity of the nanowire.

## 3. Conclusion

We have studied thermally assisted DW generation in nanowires with a perpendicular magnetic anisotropy as an alternative approach to generate DWs. The required current density to generate DWs can be effectively controlled by the proper selection of the pulse width and the constant assist field, which is applied during the current pulse. We can selectively generate a DW at a predefined location in PMA nanowires with great reproducibility, which is challenging with conventional procedures based on random nucleation sites. The proposed method is useful especially for perpendicular anisotropy materials which exhibit a high temperature dependence in coercivity. The recent technological developments in heat assisted magnetic recording which utilizes the temperature dependence of coercivity motivate our approach in DW devices. Furthermore, the pulse width requirements are lenient compared to the Oersted field generation method. The proposed method can be extended to generate any desired number of DWs in a single nanowire with relative ease compared to the Oersted field generation method.

**Acknowledgements**

This work is supported by the Singapore National Research Foundation under CRP Award No. NRF-CRP 4-2008-06.

Figure captions

Fig. 1. (a) Schematics of trilayer film stacks for sample A. (b) SEM micrograph of a 600 nm wide nanowire with 3 Hall bars placed 8 μm apart from each other. (c) Anomalous Hall measurements across Hall bars. (d) Temperature dependence of the coercivity of the nanowire.

Fig. 2. (a) Device structure with measurement schematics. The red color highlights the heated portion of the nanowire. (b) Hall bar readings show the state of magnetization.

Fig. 3. AHE measurements across $VH_1$ for the pulse with 1 ms pulse width applied across $X_1B_1$ for (a) positive and (b) negative assist field. (c,d) Respective histogram and cumulative probability of the switching processes for each polarity of assist field. (e) Switching pulse amplitude with 1 ms pulse width for various assist fields for both current polarities. (f) Switching pulse amplitude and the respective current density for 50 μs pulse width.

Fig. 4. (a) The temperature dependence of coercivity for sample B. The inset shows the Hall loop for sample B at 6 K. (b) Switching pulse amplitude with a 1 ms pulse width for various assist fields.

Fig. 5. (a) The temperature dependence of coercivity for sample C with a fit. (b) Experimental data for $J^2$ versus the pulse width with fits. (c) A Hall resistance response of domain wall depinning process at 260 K. (d) Strength of depinning fields and respective coercive fields at different temperatures.

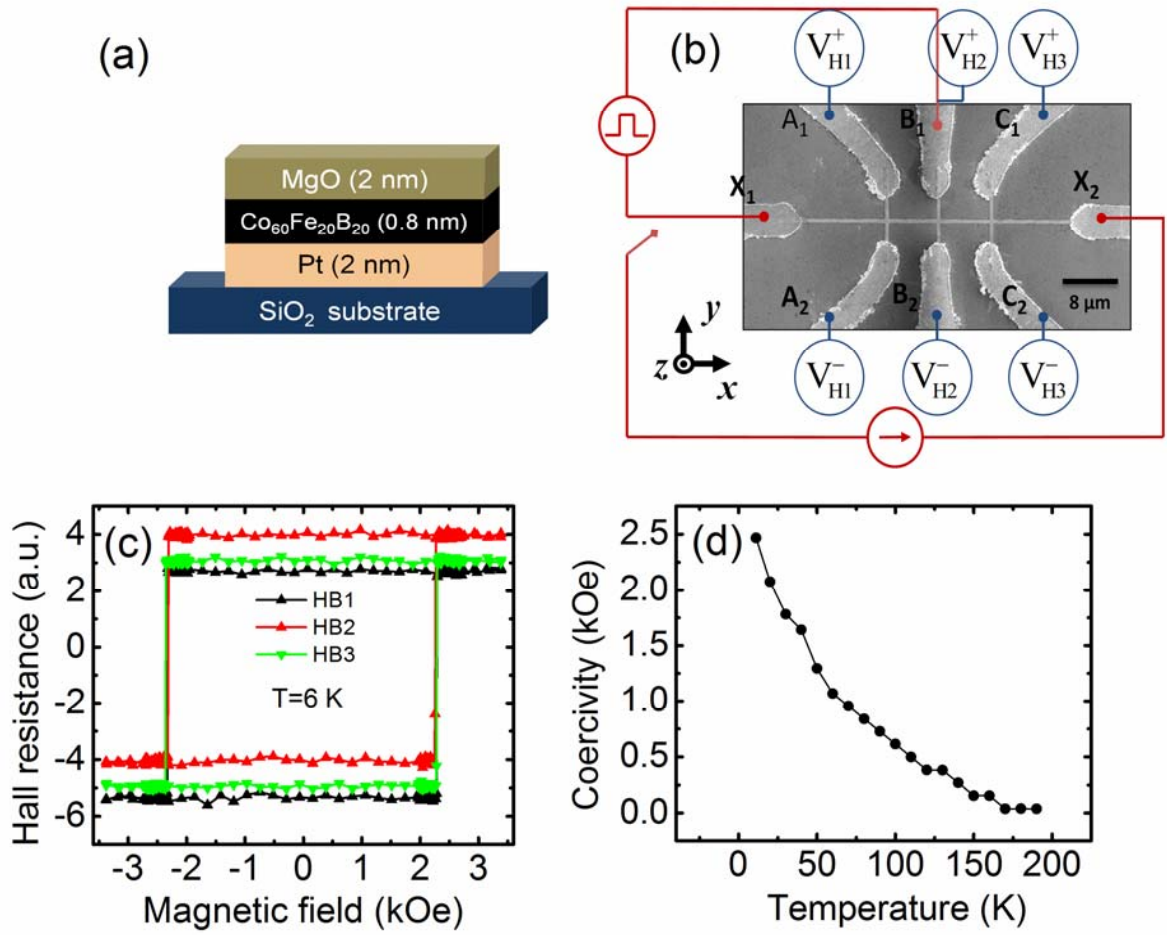

Figure 1 (Figure_1.tif)

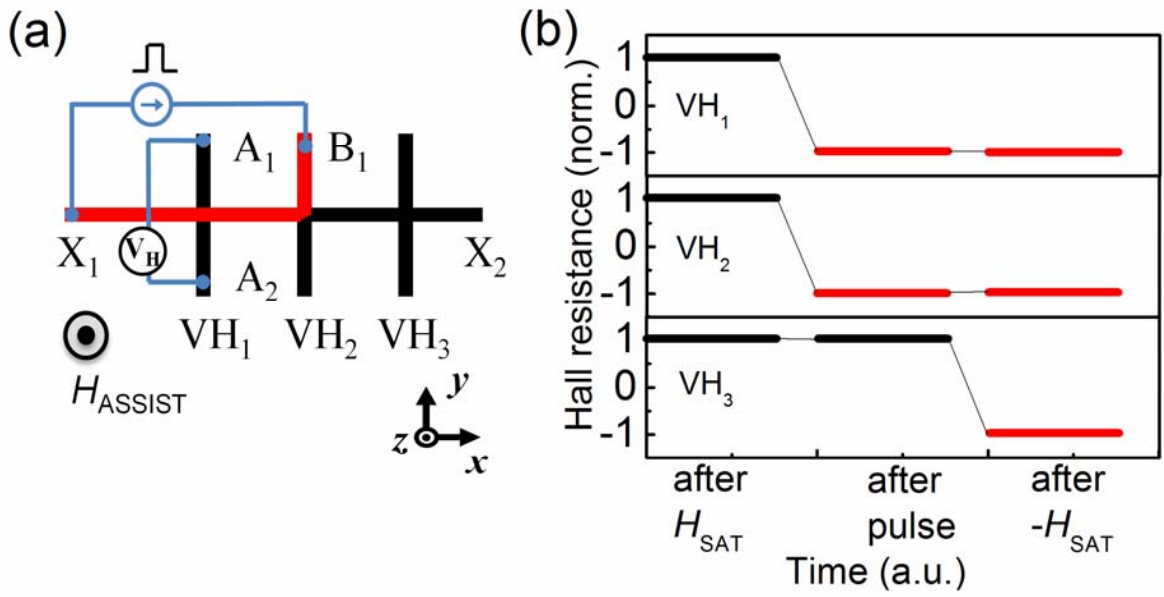

Figure 2 (Figure_2.tif)

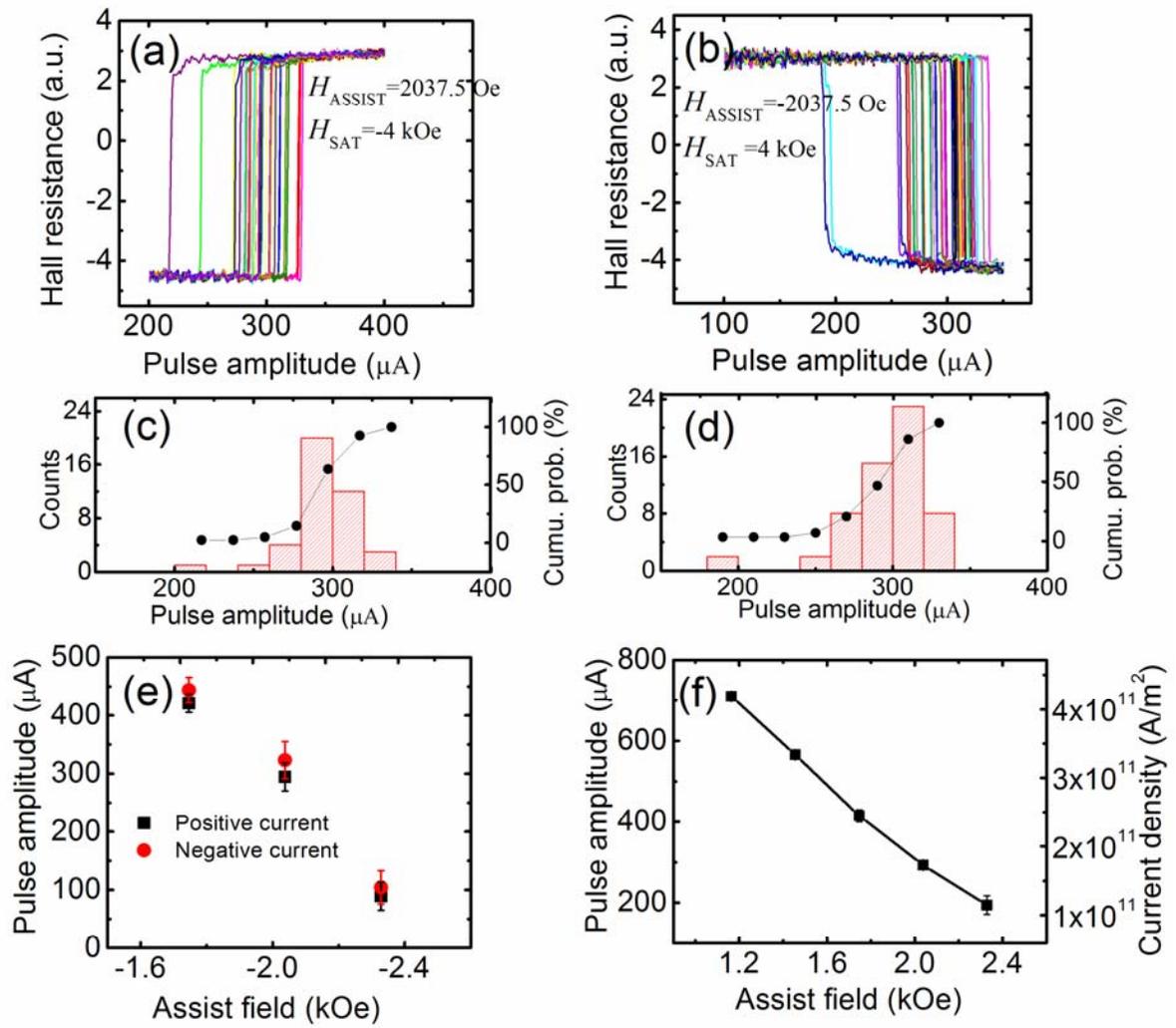

Figure 3 (Figure_3.tif)

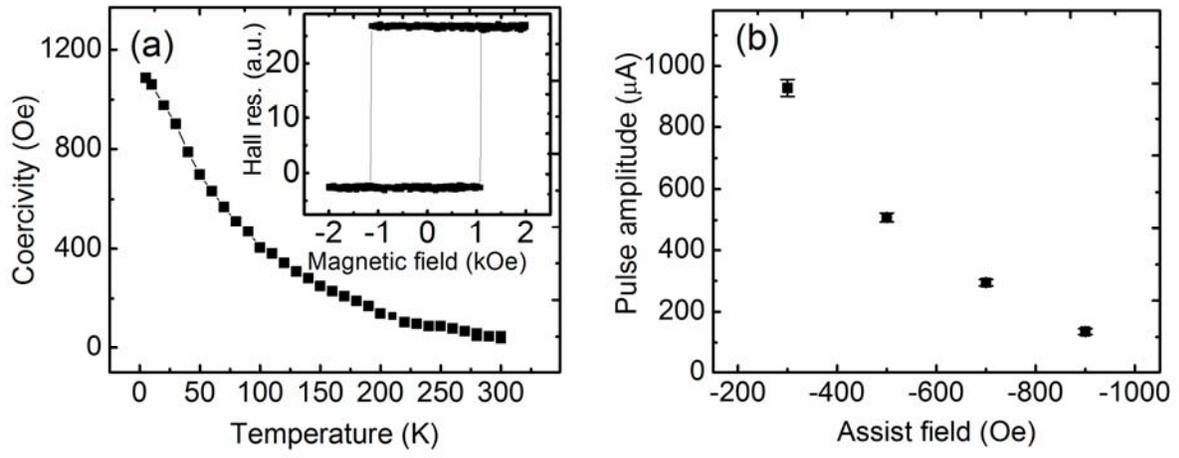

Figure 4 (Figure_4.tif)

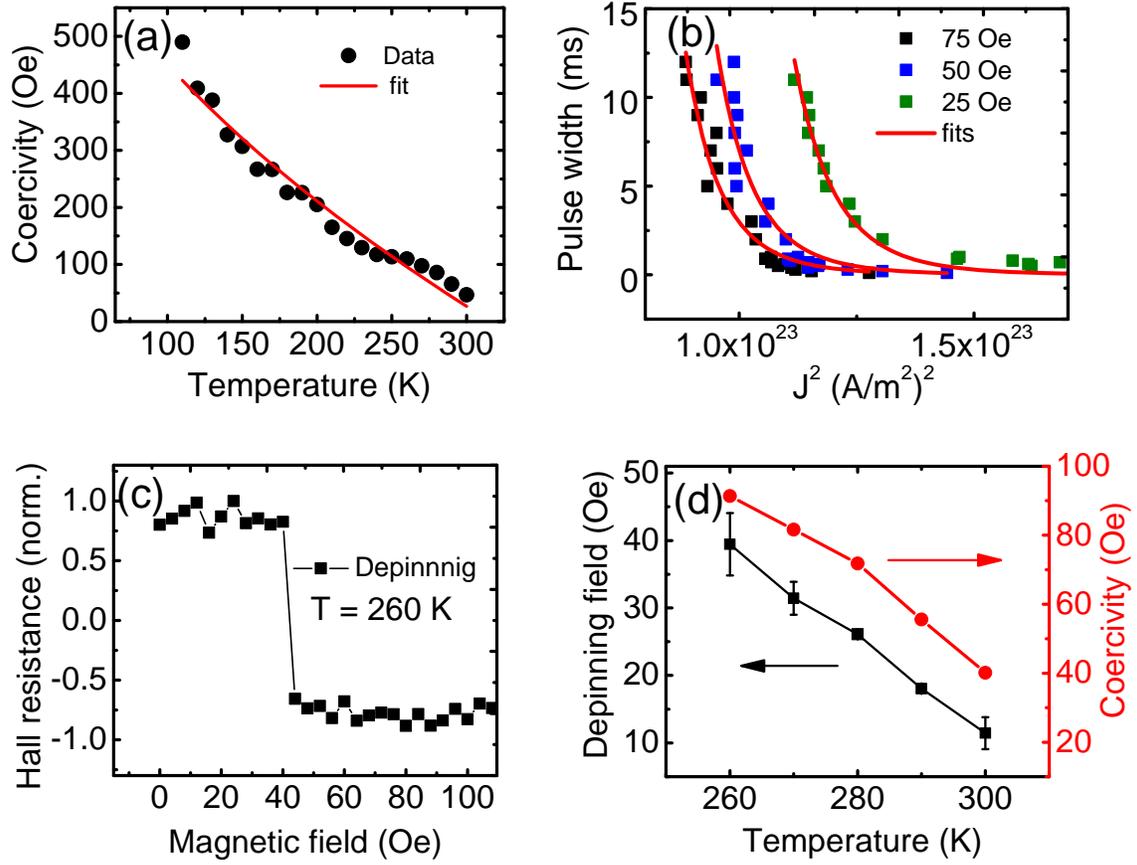

Figure 5 (Figure_5.tif)